\documentclass[letterpaper,10pt,twocolumn]{article}
\usepackage{ol2}

\usepackage{amsmath}
\usepackage{graphics}
\usepackage{epsfig}
\usepackage{amssymb}
\usepackage{color}
\usepackage{colordvi}

\begin{document}
\twocolumn[


\title{Zeno effect and switching of solitons in nonlinear couplers}
\author{
F. Kh. Abdullaev$^1$, V. V. Konotop$^1$, M. \"{O}gren$^2$, and  M. P. S\o rensen$^2$}
\affiliation{$^1$ Centro de F\'isica Te\'orica e Computacional,
Faculdade de Ci\^encias, Universidade de Lisboa, Avenida Professor
Gama Pinto~2, Lisboa 1649-003, Portugal \\
$^2$ Department of Mathematics, Technical University of Denmark, 2800 Kongens Lyngby, Denmark
}


\begin{abstract}
The Zeno effect is investigated for soliton type pulses in a nonlinear directional coupler with dissipation. The effect consists in increase of the coupler transparency with increase of the dissipative losses in one of the arms. It is shown that localized dissipation can lead to switching of solitons between the arms. Power losses accompanying the switching can be fully compensated by using a combination of dissipative and active (in particular, parity-time -symmetric) segments.
\end{abstract}
\ocis{190.0190, 060.1810, 190.5940}
]

\maketitle

\noindent

A nonlinear coupler is one of the most common and well studied optical devices (see e.g.~\cite{review}).
Two decades ago it has been realized in~\cite{Snyder}, that a coupler whose arms include dissipation and gain can greatly enrich the dynamical properties of the device.
Remarkably, the interest in a coupler with gain and loss has recently been revitalized, since it has been linked to the so-called parity-time symmetric potentials~\cite{PT1,PT2}.
It has been reported that a passive coupler, where the dissipation acts alone (without gain) may also represent a significant interest. In particular, the phenomenon of light localization was observed in a linear coupler with a dissipative arm~\cite{experiment}, where the phenomenon was attributed to the parity-time symmetry related properties. The reason for this is that in
the linear case we can invoke a trivial transformation in order to reduce the pure dissipative system to a system with gain and loss.
However, later on in~\cite{AKS}, it was argued that the increase of the transparency is intrinsic to the pure dissipative nature of the phenomenon when the dissipative losses grow. This can be observed in pure dissipative nonlinear systems as well. Even earlier this possibility was indicated in~\cite{KS} on the basis of the analogy between the mean-field dynamics of Bose-Einstein condensate and optics of Kerr media.
Moreover, it was shown in~\cite{AKS} that observation of the anomalous transparency of either a linear or a nonlinear coupler has on the one hand relation to elementary properties of a mechanical pendulum with dissipation  and on the other hand to the Zeno phenomenon.

The Zeno phenomenon,   introduced in quantum mechanics in~\cite{Zeno} and consisting in strong suppression of the decay of an unstable particle by means of permanent measurements, was  argued~\cite{KS} to have more general nature.
This connection stems from the fact that in the mean-field (quasi-classical) approximation elimination of atoms is reduced to the effect of dissipation, which is described by the equations identical to those governing the light propagation in the unidirectional coupler~\cite{AKS}. Moreover, in~\cite{AKS} it was suggested how to implement a possible experimental setup for observation of the Zeno effect in optics. In particular the decay law of continuous wave (cw) radiation and the dependence of the damping in the opposite arm of the coupler was analyzed as function of dissipation. The regimes of anomalous optical transparency of the coupler were suggested.
These studies, however, were concentrated on the steady beams, possessing no spatial-temporal structure.
This raises the question about the possibility of observing the Zeno effect for optical solitons, which is the main goal of the present Letter.

The evolution of pulses in a nonlinear directional coupler is described by the system of coupled nonlinear Schr\"{o}dinger (NLS) equations for the dimensionless fields $u_{1}(z, \tau)$ and
$u_{2}(z,\tau)$ in the two arms
\begin{eqnarray}
\label{PDE}
\begin{array}{l}
iu_{1z} +u_{1\tau\tau} + |u_1|^2u_1 +
u_2 +
i\gamma_1 u_1 =0,
\\
iu_{2z} +u_{2\tau\tau} + |u_2|^2u_2 +
u_1 +i\gamma_2 u_2 =0.
\end{array}
\end{eqnarray}

\noindent The variable $z$ measures distance along the fibers and $\tau$ measures time in a reference frame moving with the
group velocity of the optical carrier wave.
The connection is described by coupling constants, which are set to be unity, as they can be incorporated into the definition of  $\gamma_{j}$, $j=1,2$, and which are subjected to losses (or gain) described by $\gamma_{1,2}$~\cite{TWWS,Pare}

To outline the main idea, we first consider the propagation of solitons in each of the arms. Assuming that soliton shapes change gradually, we employ the Lagrangian approach 
in the coupler~\cite{Pare}, i.e., we look for solutions of (\ref{PDE}) in the form
$
u_{j}^{(s)} = A_{j}\mbox{sech}\left( \tau/a \right)e^{i\phi_j}, \label{ansatz}
$
where  the pulse parameters, i.e.  the amplitudes  $A_{1,2}$, the duration  $a$, and the phases $\phi_{1,2}$ are considered as slow functions of the propagation distance $z$.
Using this ansatz and following~\cite{Pare} we can calculate the averaged Lagrangian:
\begin{eqnarray}
\begin{array}{l} 
L= -2A_1^2a\phi_{1,z} -2A_2^2a\phi_{2,z} -\frac{2a}{3}(A_1^4 +A_2^4)
\nonumber\\
+\frac{2}{3a}(A_1^2+A_2^2) +2a A_1A_2\cos(\phi_1-\phi_2),
\end{array}
\end{eqnarray}
where the amplitudes, the relative phase $\phi=\phi_1-\phi_2$, and the pulse duration  are considered as dynamical variables. The associated Lagrange equations lead  to the following dynamical system (see e.g.~\cite{Pare} for the technical details):
\begin{subequations}
\label{mODE}
\begin{eqnarray}
F_{z} &=& -\gamma (1-F^2)+ 2   \sqrt{1-F^2}\sin(\phi)   , \label{mODE1}\\
\phi_z &=& \delta F Q -2 \frac{F}{\sqrt{1-F^2}} \cos(\phi) ,\label{mODE3} \\
Q_{z} &=& - \gamma_1 Q(1+F)-\gamma_2 Q(1-F). \label{mODE2}
\end{eqnarray}
\end{subequations}
Here $F=(P_1-P_2)/(P_1+P_2)$ characterizes the relative
distribution of the power $P_{1,2}\left(z\right)=\int_{-\infty}^\infty|u_{1,2}\left(z,\tau\right)|^2d\tau$ between the arms and $Q=(P_1+P_2)/P_0$ is the total power normalized to the input power $P_0$ at the coupler input $Q(z=0)=1$
and $ \delta= P_0/3a$.
Notice that in the obtained approximation the pulse width $a$ is constant 
which explains why Eqs.~(\ref{mODE}) are identical to those obtained and studied in~\cite{AKS} for the coupler operating in the stationary regime.
This also means that one can expect various types of dissipative dynamics with solitons, similar to those observed in the stationary regime.

\begin{figure}[tbp]
\begin{center} 
\includegraphics[height=3.4cm]{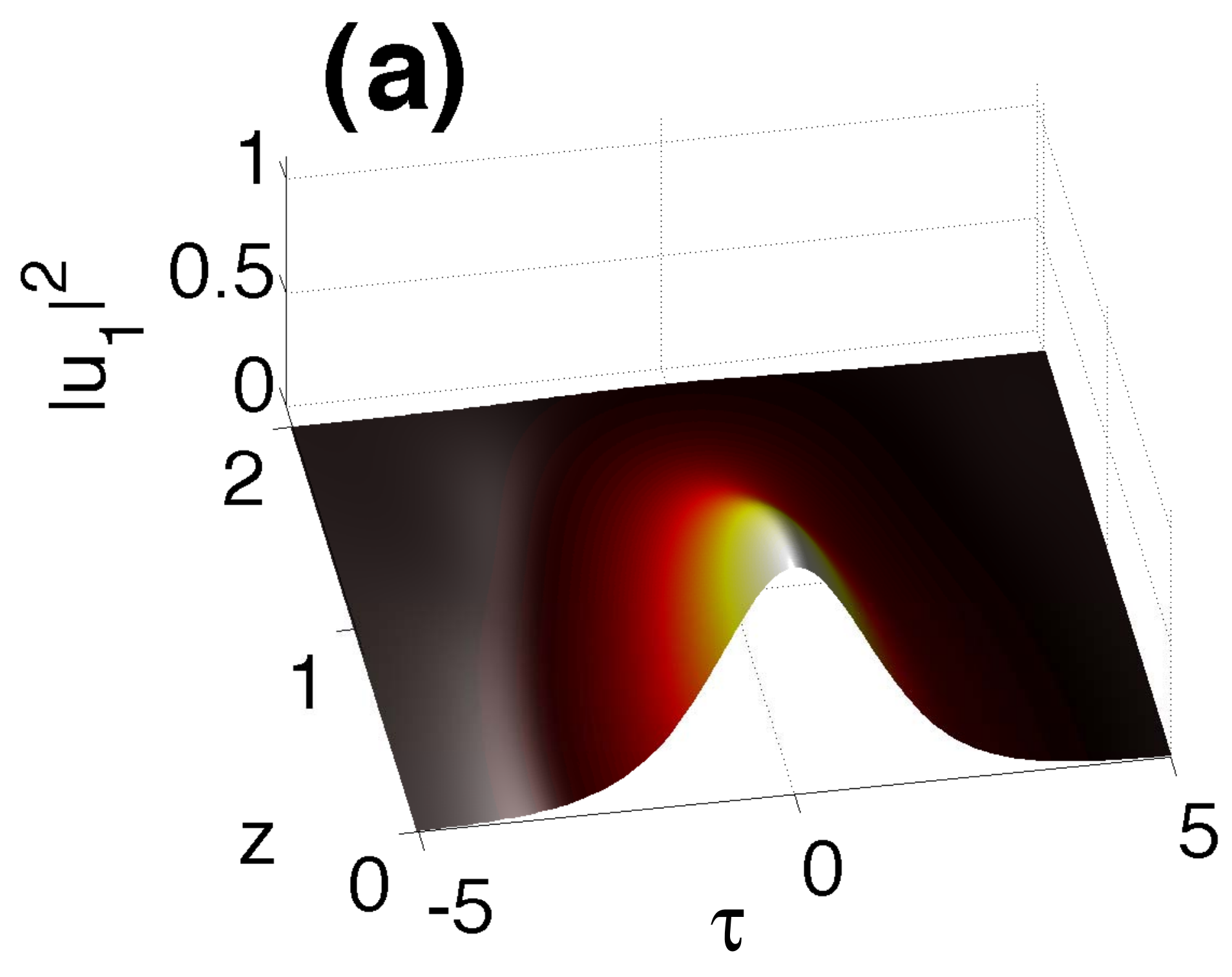}~~
\includegraphics[height=3.4cm]{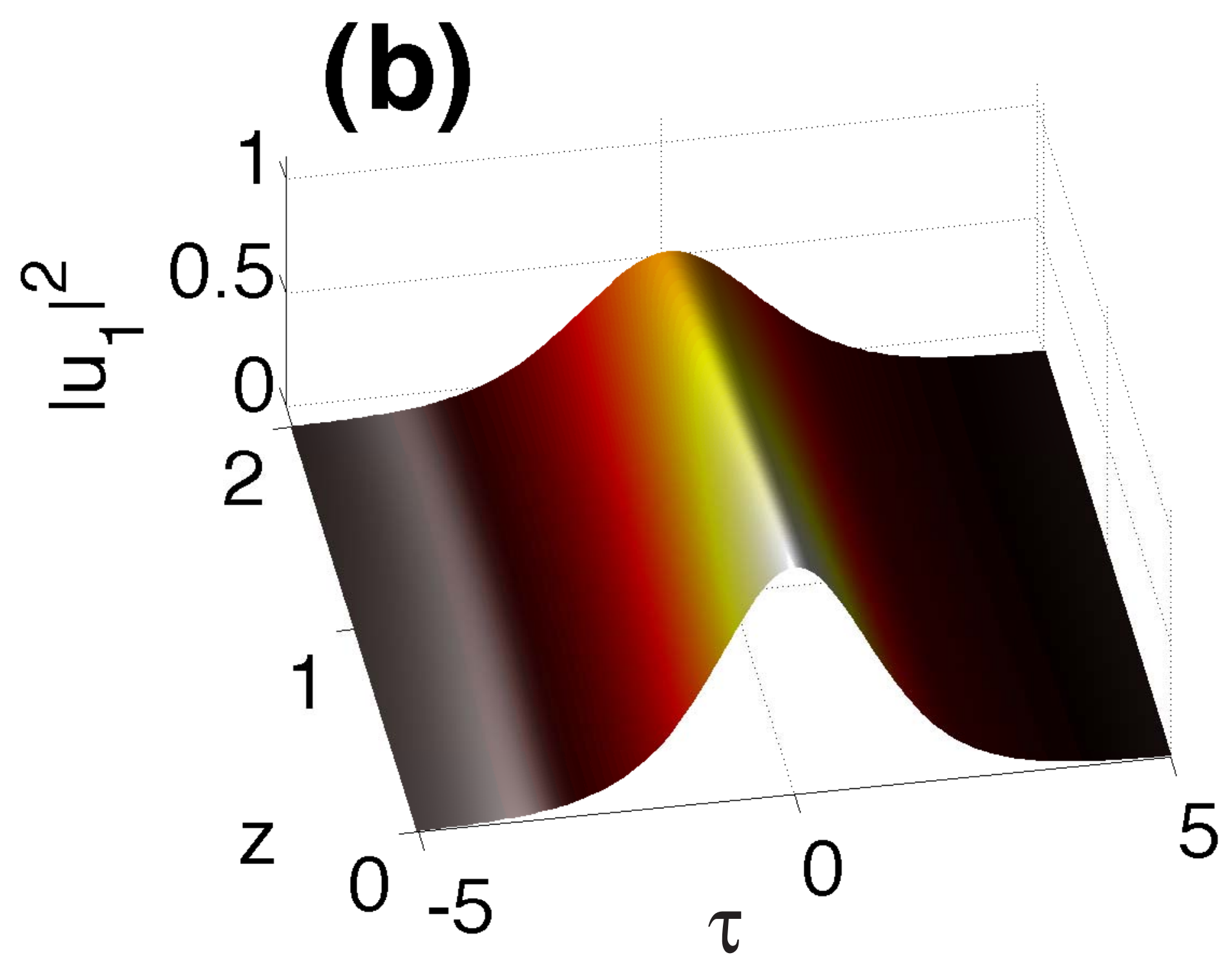}
\end{center}
\caption{(Color online) Observation of a Zeno soliton in component $u_1$ for large enough damping  ($\gamma_2$) in component $u_2$ for the input pulse of the form $u_1 (z=0)=\mbox{sech}\left(\tau /\sqrt{2}\right)$, $u_2(z=0)=0$.
(a)~Abrupt decay of the field $u_1$ for $\gamma_2=1$. (b)~A Zeno soliton in $u_1$ for $\gamma_2=10$.}
\label{figure1}
\end{figure}

Now we turn to the direct numerical simulations of Eqs.~(\ref{PDE}).
First, in order to emphasize the main effects, we will assume that one of the arms ($j= 1$) is transparent, i.e., $\gamma_1=0$, and study the dynamical regimes depending on the losses $\gamma_2$ in the second arm.
In Fig.~\ref{figure1}  we illustrate our main result.
Here we observe that applying the input pulse to the transparent arm for a relatively weak dissipation, the output power in both arms display exponential decay [panel (a)], while an increase of the losses by a factor 10, one strongly suppresses the power loss in the transparent arm, giving rise to a "Zeno soliton" in component $u_1$ at the output [panel (b)].

The two cases in Fig.~\ref{figure1} correspond to the input power below the threshold of switching of the conservative coupler (i.e. (\ref{mODE}) with $\gamma_{1,2}=0$), which for the stationary coupler is $A_{cr}^2=6$~\cite{review}, while for the coupler operating in the solitonic regime it is $A_{cr}^2 \approx 6.7$~\cite{review,Pare}.

\begin{figure}[tbp]
\begin{center} 
\includegraphics[height=3.25cm]{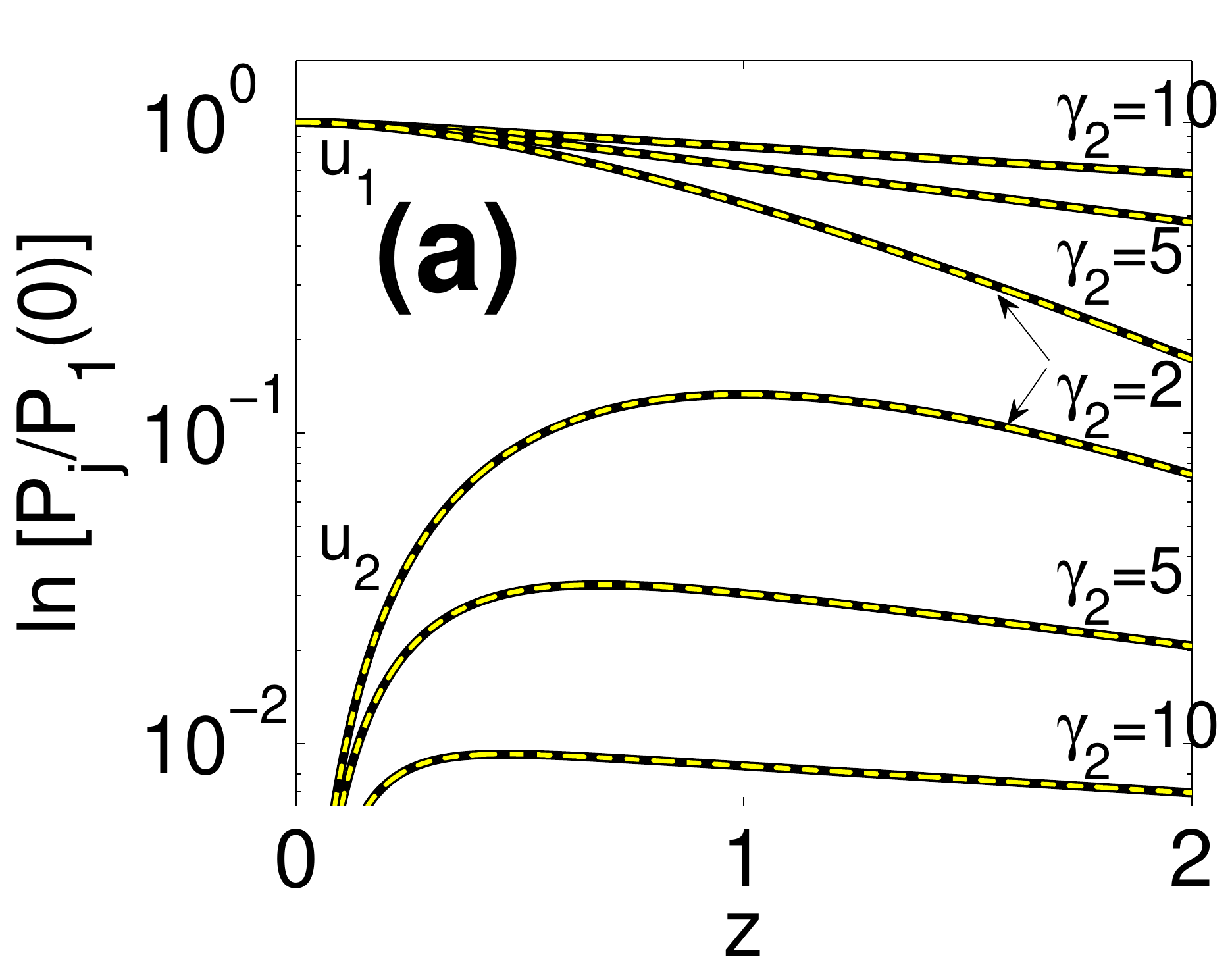}~~
\includegraphics[height=3.1cm]{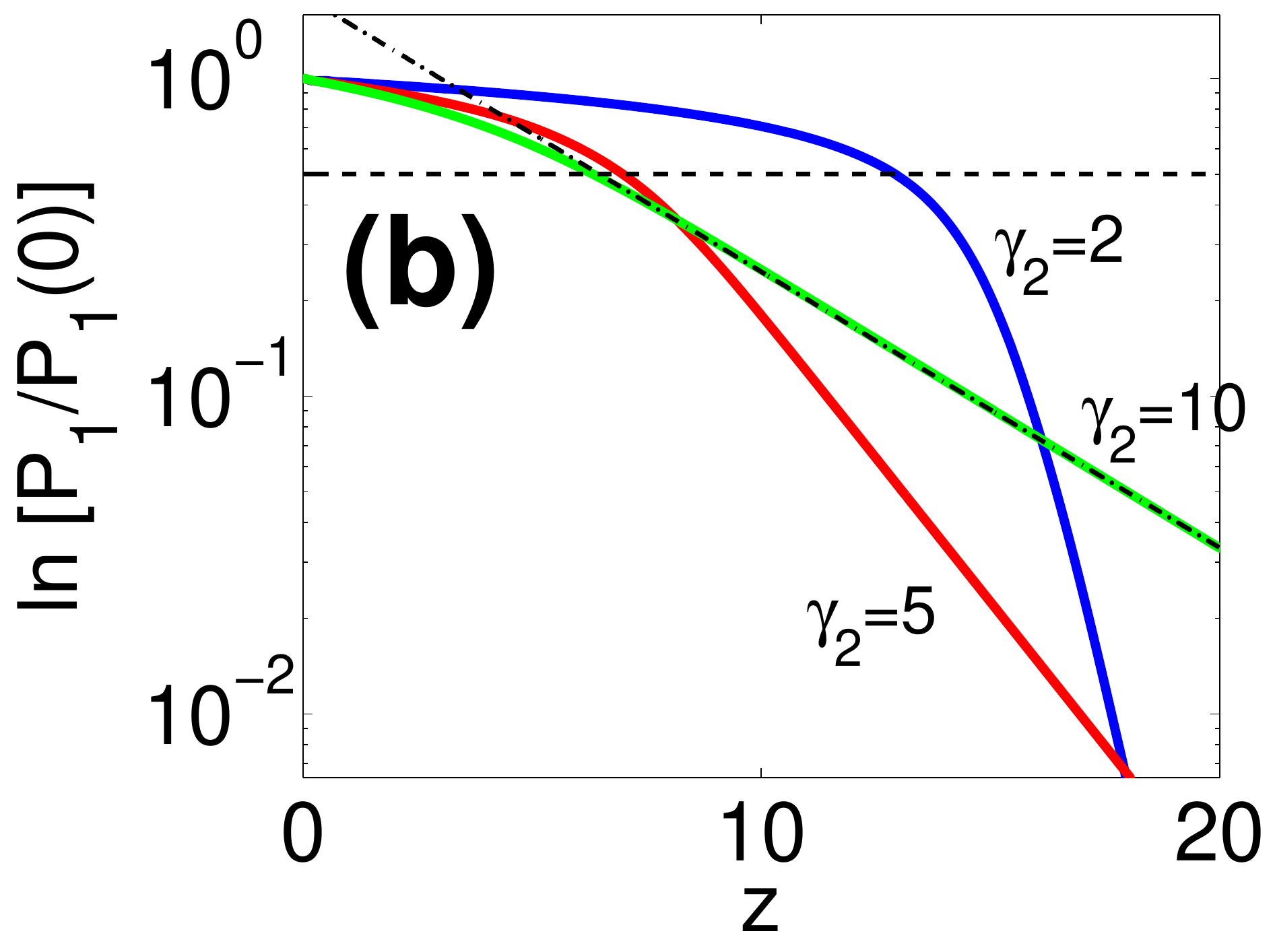}
\end{center}
\caption{(Color online) Distribution and decay of powers $P_j$.
 (a)~The exponential decay in the first arm and local energy transfer between the arms for the initial data as in Fig.~\ref{figure1} (i.e. below the threshold, $A_1^2< 6.7 $). 
Black solid curves shows the full simulation of Eq.~(\ref{PDE}), while yellow (gray) dashed curves shows the approximate  results obtained from (\ref{mODE}).
(b)~The same as in panel (a) but for the input amplitude $A_1=5$, i.e. above the  threshold,  $A_1^2>6.7$.
The horizontal dashed line shows the normalized critical threshold value $A_{cr} \simeq \sqrt{6.7 }$  (i.e. the value $\sqrt{6.7}/5\simeq 0.5$). 
The sloop $-2/\gamma_2$ is illustrated with the dashed-dotted line for $\gamma_2=10$.} 
\label{figure2}
\end{figure}

Below the threshold the emergence of a Zeno soliton occurs independently of the input power $P_1\left( 0 \right)$.
However, for  input powers  above the threshold the situation is changed. There appear two very different regimes of the soliton decay, as is shown in
Fig.~\ref{figure2}. 
Starting with the undercritical input power   we clearly observe the Zeno phenomenon, expressed in the decrease of the decay exponent [Fig.~\ref{figure2}(a)]. It is remarkably well   described by the law  $P_1/P_1(0) \simeq \exp \left( -2
z/\gamma_2 \right)$. 
This law follows from the system (\ref{mODE}) of the adiabatic  change of the soliton parameters, and hence its physics is the same as described in~\cite{AKS} for the stationary coupler
[we emphasize the very accurate approximation given by the system (\ref{mODE}) here]. The field amplitude in the second arm, which is zero at the input, achieves its maximum at short distances (being always a few orders below the field amplitude of the first component) and after that rapidly decays. We also notice that energy transfer between the transparent and lossy arms is strongly suppressed by the losses.
If the input power is well above the critical value, the situation is first inverted, and at small distances one observes a "standard" decay exponent proportional to $\gamma_2$.
However, as soon as the pulse power has decreased to some critical value, well estimated by $A_{cr}\approx 6.7$ [horizontal dashed line in Fig.~\ref{figure2}(b)], a change of the exponential decay, i.e., the Zeno effect, is observed.

\begin{figure}[tbp]
\begin{center} 
\includegraphics[height=3.2cm]{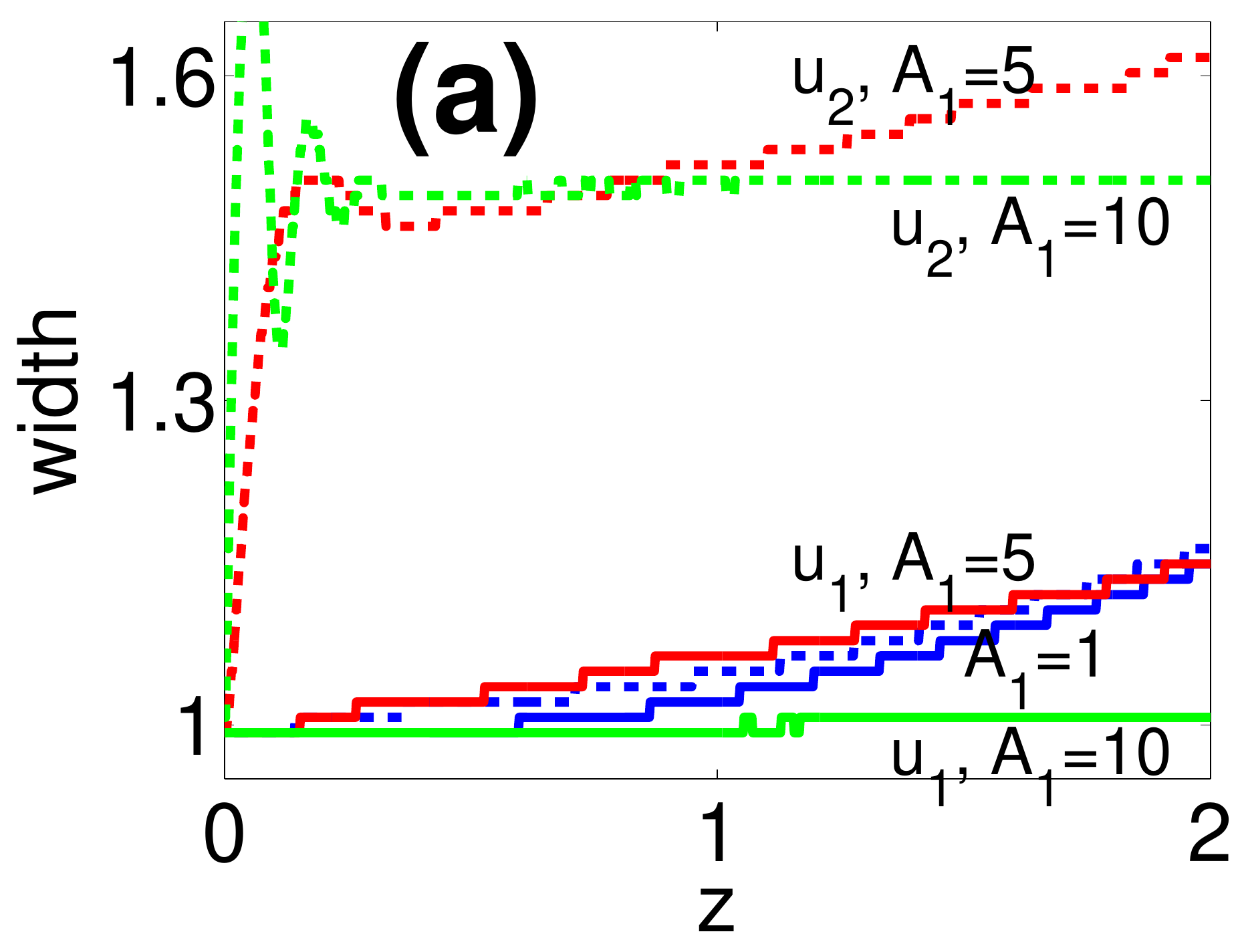}~~
\includegraphics[height=3.2cm]{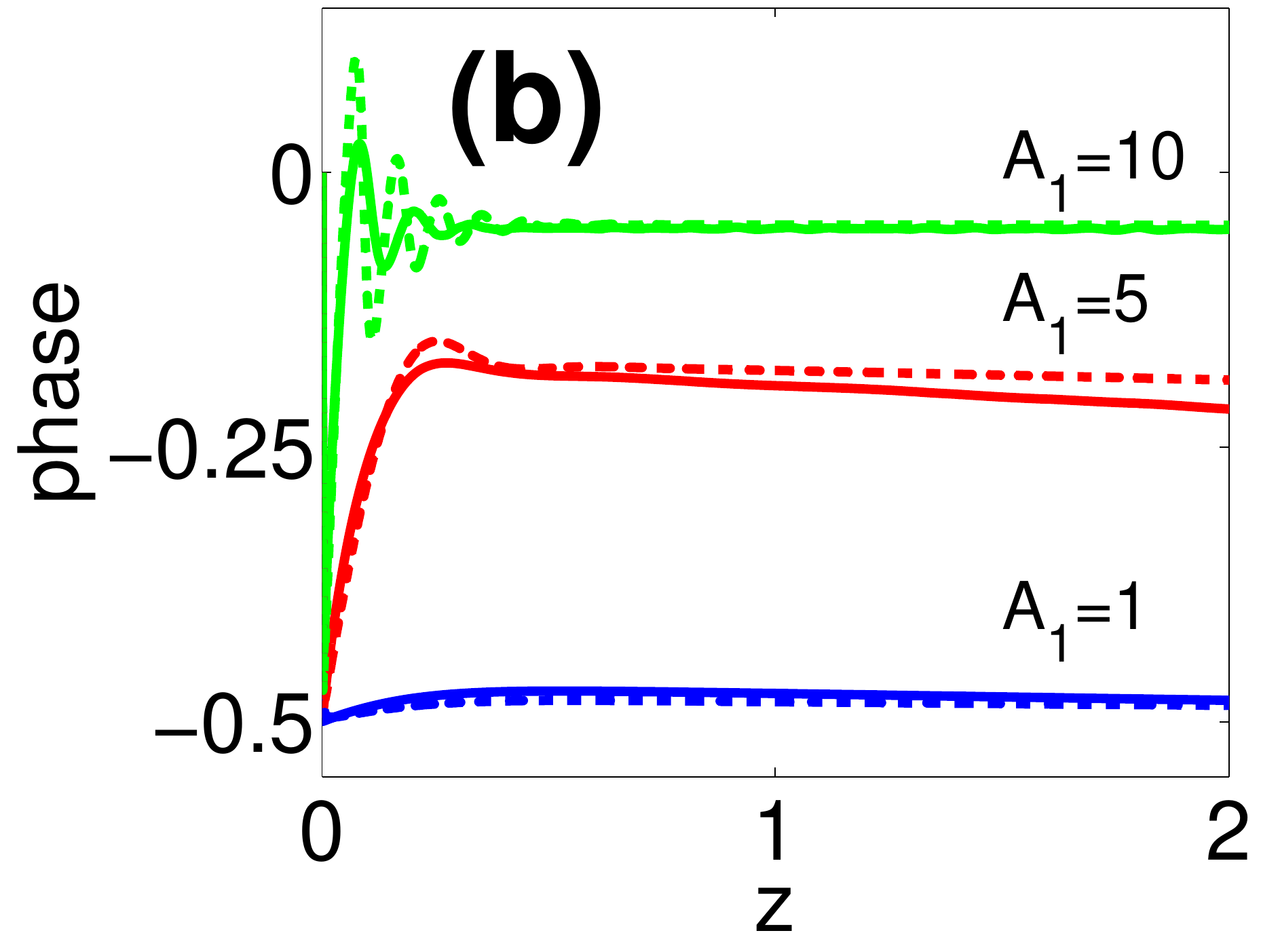}
\end{center}
\vspace{-0.5cm}
\caption{(Color online) (a) The half-width at the half maximum {\it vs} propagation distance for different input amplitudes $A_1$ ($a=\sqrt{2}/A_1$) and a fixed $\gamma_2=10$.
Solid (dashed) curves are for arm $j=1 \: (2)$.
(b) The relative central phase $\phi(z,\tau=0)$ obtained from (\ref{PDE}) (solid curves) and from (\ref{mODE}) (dashed curves), normalized with $\pi$.}
\label{figure3}
\end{figure}

To test the accuracy of the predictions given by (\ref{mODE}), we performed
numerical comparisons of the half-width at the half-maximum of the solution (which for $u_1^{(s)}(\tau)$ is given by $a\,\mbox{arccosh}(\sqrt{2})$  used in Fig.~\ref{figure3}(a) to normalize the width), and of the relative central phase $\phi(\tau=0, z)$ with the corresponding values of $\phi$ obtained from (\ref{mODE}), see Fig.~\ref{figure3}. 
The peculiarity is that the width of the wavepacket in the lossy arm is about 1.5 times bigger than the width of the input soliton, which can be explained by the dominating dispersion in the second arm due to low intensity of the pulse.

\begin{figure}[tbp]
\begin{center} 
\includegraphics[height=3.2cm]{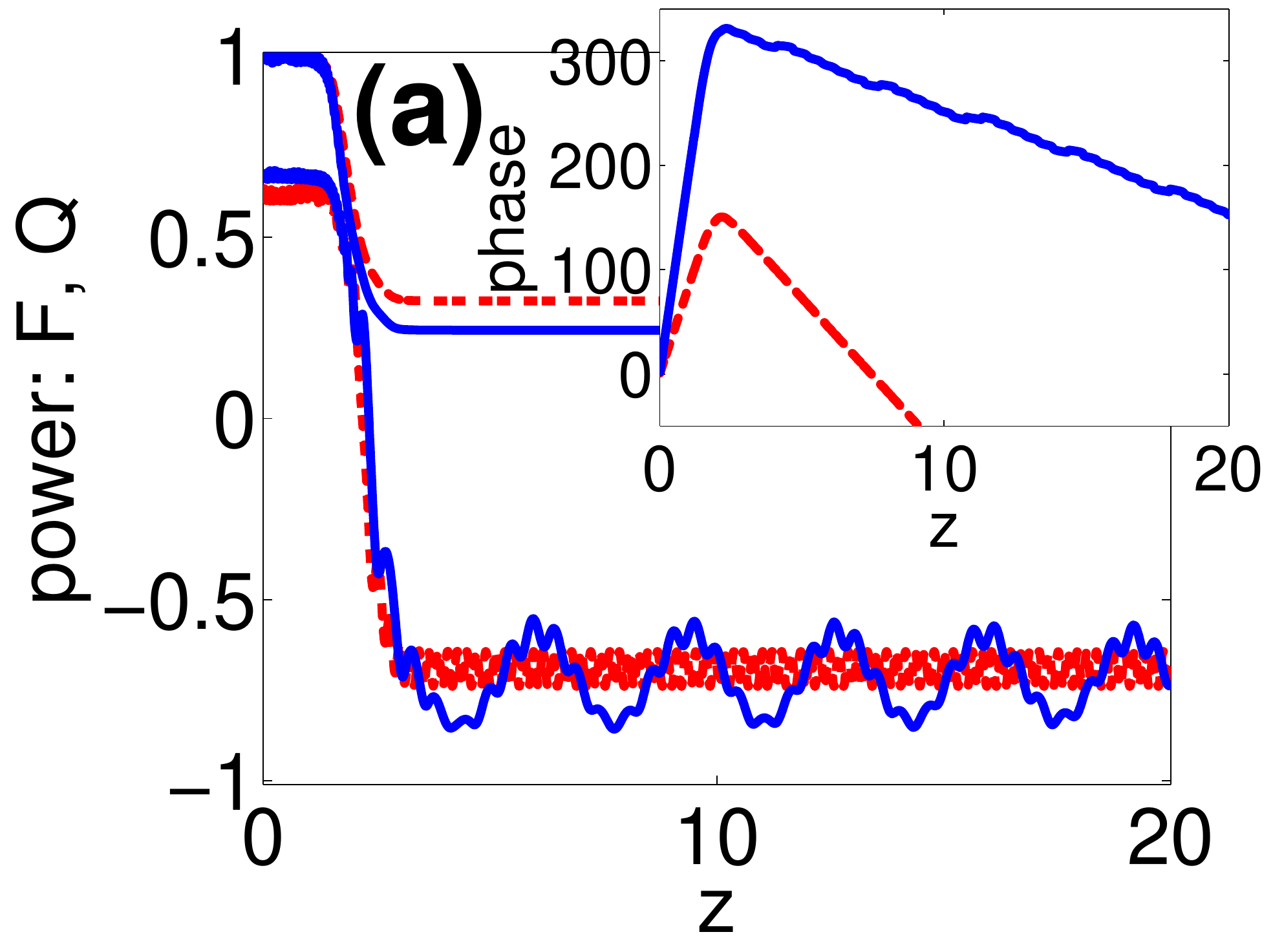}~~
\includegraphics[height=3.25cm]{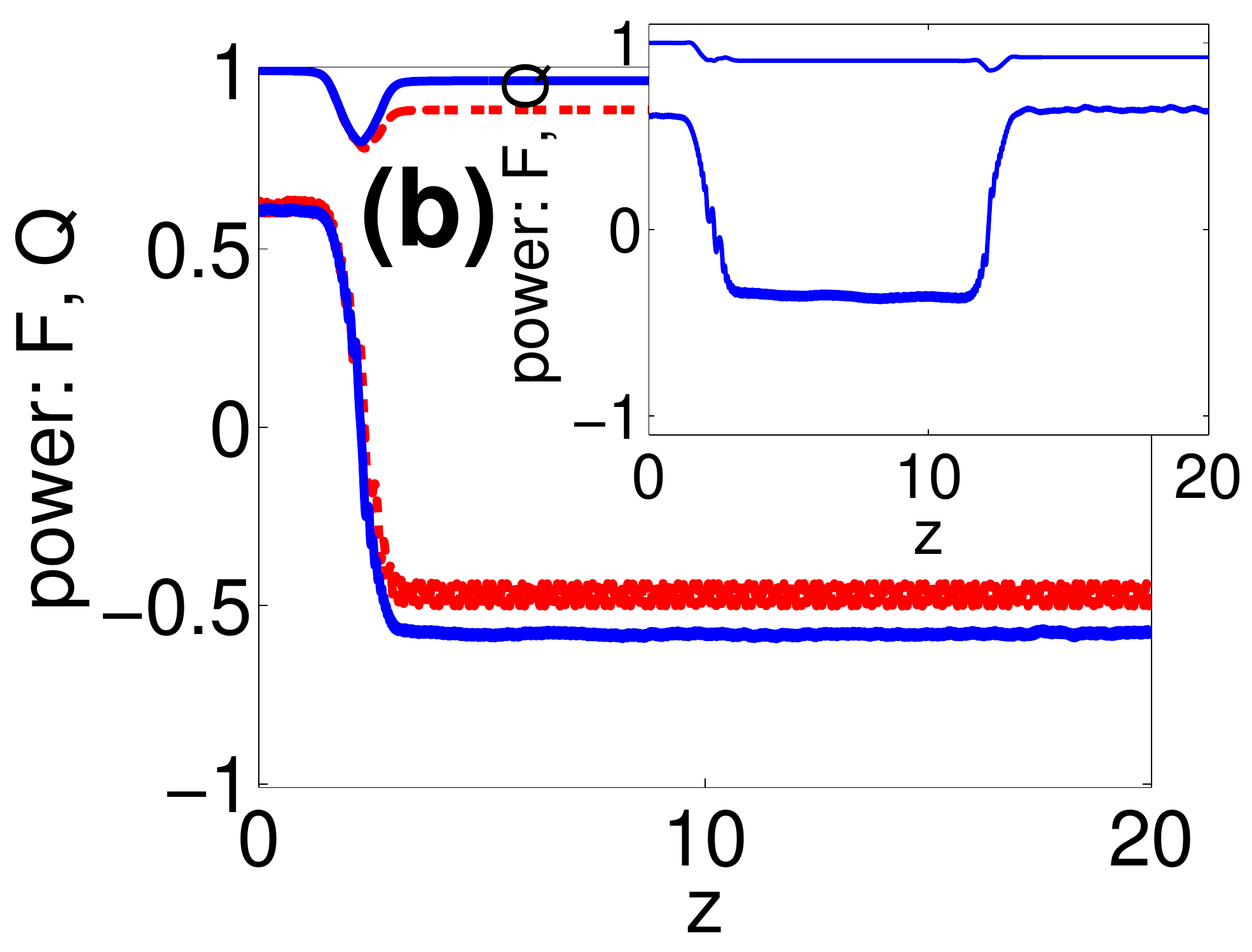}
\end{center}
\vspace{-0.5cm}
\caption{(Color online) Switching between the arms induced by localized defects.
In (a) only a dissipative segment $\gamma_1(z)=\Gamma \left( \arctan [5(z-z_a)] -  \arctan [5(z-z_b)] \right)^2$
with $\Gamma=0.135$, $z_a=1.5$, and $z_b=3.0$ in the first arm
is included.
The inset shows the relative phase.
The solid (blue) and dashed (red) curves show the simulations of the models in (\ref{PDE}) and (\ref{mODE}), respectively.
In (b) the simultaneous effect of the dissipative and active segments, representing a ${\cal PT}$-like structure, i.e.  $\gamma_2(z)= - \gamma_1(z)$, with $\gamma_1(z)$ as in panel (a) but with $\Gamma= 0.065$.
For the inset of (b) we have added a segment of gain/dissipation at a $\Delta z \sim 10$ distance, and used engineered parameters in order to keep $Q(z) \approx 1$.
In both (a) and (b) we use $u_1(z=0)=20\, \textnormal{sech}(10\sqrt{2} \tau ) $ and  $u_2(z=0)=5\, \textnormal{sech}(5  \tau /\sqrt{2} ) $, i.e. initially $F(0)=3/5$.  }
\label{figure4}
\end{figure}

It has been shown in~\cite{KS} that rich possibilities of the control of the stationary modes in   a two-mode (dimer)  system can be achieved by using localized dissipation. In our case this scenario can be realized by using coupler arms, which are transparent everywhere except at localized dissipative segments. Now we turn to the dynamics of solitons in such systems focusing on the possibilities of implementing different switching regimes. 
%
The first simple illustration of the soliton switching is shown in Fig.~\ref{figure4}(a).
At the input of the coupler both arms are transparent and the coupler is operating in the "soliton locked"  regime (i.e. above the threshold).
Without dissipation the main power would propagate in the first arm.
If however, along the propagation distance in the first arm, a short dissipative segment is included,  one observes  switching of the power  to the second arm ($F<0$).
This process is accompanied by the loss of more than 50\% of the total power and with the change of the "direction" of rotation of the relative phase [the inset of  Fig.~\ref{figure4}(a)].
The switching phenomena for solitons seen in Fig.~\ref{figure4} can be qualitativelly, but not quantitatively, reproduced by the approximate ODE model~(\ref{mODE}), which only have one parameter for the width $a \equiv (a_1+a_2)/2$.

The problems of losses  can be completely resolved by using simultaneously dissipative and active gain segments.
In Fig.~\ref{figure4}(b) we use a parity-time (${\cal PT}$) like segment,   $\gamma_2(z)=-\gamma_1(z)$,  and observe that the total power do not decrease dramatically anymore. We emphasize, however, that in spite of the similarity of our model with ${\cal PT}$ structures
the effect can be observed (and  even optimized) using the engineering of the dissipative and gain defect, which do not have well defined symmetry (i.e. having different lengths or amplitudes).
Finally, in the inset of Fig.~\ref{figure4}(b), we have engineered the amplitudes ($\Gamma$:s) and the translations ($z_{a,b}$:s) of the first dissipation/gain segment,
and in addition added a second segment of gain/dissipation in order to switch back, and with the total power being kept close to $Q(z) \approx 1$.

To conclude we have reported the possibility of observation of the Zeno effect with solitons in a coupler with one transparent and one lossy arm. The effect consists in the increase of the total transparency of the coupler when the losses of one of the arms are increased. We have also shown that using dissipative segments one can perform efficient management of solitons, in particular switching them at will. Since the intensity losses are inevitable in the described systems, we have suggested simultaneous use of the dissipative and gain segments, 
in order to perform lossless manipulations of the solitons.

  \medskip

FKA and VVK were supported by  the FCT (Portugal) under the   PEst-OE/FIS/UI0618/2011 and by the grant PIIF-GA-2009-236099 (NOMATOS) within the 7th European Community Framework Programme.
FKA acknowledge appointment as Otto M{\o}nsted Guest Professor during visit to DTU and M\"{O} is supported by the H.C. {\O}rsted post.doc.
programme.


\begin{thebibliography}{99}

\bibitem{review} M. Romagnioli, S. Trillo, and S. Wabnitz, Opt. Quant. Electron. {\bf 24} S1237-S1267 (1992).

\bibitem{Snyder} Y. Chen, A. W. Snyder, and D. N. Pain, IEEE J. Quant. Electron. {\bf 28}, 239-245 (1992).

\bibitem{PT1} H. Ramezani, T. Kottos, R. El-Ganainy, and D. N. Christodoulides,  Phys. Rev. A {\bf 82}, 043803 (2010).

\bibitem{PT2} A. A. Sukhorukov, Zhiyong Xu, and Y. S. Kivshar,  Phys. Rev. A {\bf 82}, 043818 (2010).

\bibitem{experiment} A. Guo, G. J. Salamo, D. Duchesne, R. Morandotti, M. Volatier-Ravat, V. Aimez, G. A. Siviloglou, and D. N. Christodoulides, Phys. Rev. Lett. {\bf 103}, 093902 (2009).

\bibitem{AKS}
F. Kh. Abdullaev, V. V.  Konotop and V. S.  Shchesnovich, Phys. Rev. A {\bf 83}, 043811 (2011).

\bibitem{KS}
V. S.  Shchesnovich and V. V. Konotop,  Phys. Rev. A {\bf 81}, 053611 (2010).

\bibitem{Zeno} B. Misra and E. C. G. Sudarshan, J. Math. Phys. Sci. {\bf 18}, 756 (1977).

\bibitem{TWWS} S. Trillo, S. Wabnitz, E. M. Wright, and G. I. Stegeman,  Opt. Lett., {\bf 13}, 672 (1988).

\bibitem{Pare}
C. Pare and  M. Florjanczyk, Phys. Rev. A {\bf 41}, 6287-6295  (1990).

\end{thebibliography}
\end{document}